# Sluggish and Chemically-Biased Interstitial Diffusion in Concentrated Solid Solution Alloys: Mechanisms and Methods

Biao Xu [a b], Haijun Fu [a], Shasha Huang [a], Shihua Ma [a], Yaoxu Xiong [a], Jun Zhang [a], Xuepeng Xiang [a], Wenyu Lu [a], Ji-Jung Kai [a b d*], Shijun Zhao [a c d*]

[a] Department of Mechanical Engineering, City University of Hong Kong, Hong Kong 999077, China

[b] Center for Advanced Nuclear Safety and Sustainable Development, City University of Hong Kong, Hong Kong 999077, China

[c] City University of Hong Kong Shenzhen Research Institute, Shenzhen 518057, China

[d] Hong Kong Institute for Advanced Study, City University of Hong Kong, Hong Kong 999077, China

Email: jijkai@cityu.edu.hk (Ji-Jung Kai) and shijzhao@cityu.edu.hk (Shijun Zhao)

# Abstract

Interstitial diffusion is a pivotal process that governs the phase stability and irradiation response of materials in non-equilibrium conditions. In this work, we study sluggish and chemically-biased interstitial diffusion in Fe-Ni concentrated solid solution alloys (CSAs) by combining machine learning (ML) and kinetic Monte Carlo (kMC), where ML is used to accurately and efficiently predict the migration energy barriers on-the-fly. The ML-kMC reproduces the diffusivity that was reported by molecular dynamics results at high temperatures. With this powerful tool, we find that the observed sluggish diffusion and the "Ni-Ni-Ni"-biased diffusion in Fe-Ni alloys are ascribed to a unique "Barrier Lock" mechanism, whereas the "Fe-Fe-Fe"-biased diffusion is influenced by a "Component Dominance" mechanism. Inspired by the mentioned mechanisms, a practical AvgS-kMC method is proposed for conveniently and swiftly determining interstitial-mediated diffusivity by only relying on the mean energy barriers of migration patterns. Combining the AvgS-kMC with the differential evolutionary algorithm, an inverse design strategy for optimizing sluggish diffusion properties is applied to emphasize the crucial role of favorable migration patterns.

# Introduction

Concentrated solid-solution alloys (CSAs), a term encompassing alloys composed of multiple principal elements like high-entropy alloys (HEAs), have garnered growing research attention due to their multifaceted properties, such as superior fracture resistance [1], tensile strength [2,3], corrosion resistance [4,5], thermal stability [6-8], and irradiation tolerance [9-13]. The prominence of these properties is closely linked to the phenomenon of sluggish diffusion and chemically-biased diffusion within CSAs. Sluggish diffusion [14-16], characterized by a slower diffusivity in CSAs with specific compositions compared to any constitutive pure phases, extends the material's service life before encountering instability or undesired microstructural degradation. On the other hand, chemically-biased diffusion denotes a scenario in which atomic transportation exhibits a distinct affinity for one of the constituent species within CSAs. This phenomenon serves as a valuable guide for designing desirable alloys by fine-tuning the compositions that drive biased diffusion. While both these diffusion mechanisms have been extensively studied, particularly in the context of vacancy-mediated self-diffusion in CSAs [17-19], the understanding of sluggish and chemically-biased diffusion for interstitial-mediated diffusion remains insufficient.

The diffusion of vacancies and interstitials plays a pivotal role in driving structural transformations under non-equilibrium conditions, such as deformation and irradiation. Notably, interstitials typically exhibit diffusivity orders of magnitude higher than that of vacancies. Consequently, comprehending the interstitial diffusion mechanism within CSAs is of paramount importance for a comprehensive understanding and accurate prediction of changes in their microstructure. Experimental observations have confirmed that the extent of damage, as evidenced by the positioning of interstitial dislocation loops within Fe-Ni CSAs, is notably reduced compared to pure metals [20]. This compellingly supports the presence of sluggish interstitial-mediated diffusion in CSAs. Additionally, significant enrichment and depletion of constituent elements in CSAs suggest chemically-biased diffusion and segregation bias [21]. Theoretically, interstitial diffusion in CSAs has been explored using methods such as *ab initio* molecular dynamics (AIMD) [22], classical molecular dynamics (MD) [19,22-24], and kinetic Monte Carlo (kMC) [19,24,25]. The observed phenomena of sluggish diffusion and chemically-biased diffusion, as established in theoretical studies, are in alignment with experimental findings [19,20]. Despite substantial progress, the role of chemical disorder, the most critical CSA characteristic, remains elusive, particularly regarding its impact on diffusion.

Unraveling the effects of chemical disorder on diffusion mechanisms presents a significant challenge for traditional theoretical methodologies, with the mentioned techniques facing inherent limitations. The AIMD approach typically necessitates limiting the simulation size to a few hundred atoms and timescales to sub-nanoseconds due to the substantial computational demands. This constrained system size often results in an oversimplification of the intricate atomic environments associated with disorder in CSAs, and the short timescales frequently lead to diffusion properties with inadequate statistical significance, particularly in the context of CSAs, where diffusion tends to be sluggish. While classical MD in conjunction with interatomic potentials can offer relatively dependable diffusion modeling within a practical simulation box size and over timescales spanning microseconds [17,19,24], it struggles to provide substantial insights into the long-range diffusion processes and may result in an incomplete comprehension of the diffusion mechanisms [26]. The kMC method is capable of simulating diffusion processes up to a timescale of seconds, yet it has its limitations when it comes to precisely characterizing the diffusion mechanisms within CSAs. The challenge lies in acquiring the transition rates, denoted by $\Gamma = \sum_i \Gamma_0 \exp\left(-\frac{E_b^i}{k_B T}\right)$, which are essential inputs for the kMC algorithm. In this formula, $\Gamma_0$ represents the attempted frequency, $E_b^i$ denotes the transition energy barrier for possible jump path $i$, $k_B$ is the Boltzmann constant, and $T$ signifies the temperature. Calculating the transition rate heavily relies on the most critical parameter $E_b^i$, which is typically determined through time-expensive nudged elastic band (NEB) calculations [27,28]. However, the diverse local atomic environment combinations within CSAs imply that NEB calculations are essential for every potential jump at each step, especially when the diffusing interstitial changes its position. As a result, directly implementing cumbersome NEB calculations within the kMC algorithm becomes impractical. In fact, previously proposed kMC approaches in CSAs have relied on substantial approximations to account for the chemical-disorder-dependent energy barriers. For instance, the variable energy barriers of interstitials in Fe-Ni CSAs have been simplistically represented by fixed values from their pure phases, potentially yielding an unreliable comprehension of the diffusion mechanism [24,25].

In our previous work [29], we demonstrated the effectiveness of the machine learning (ML) and kMC combination in uncovering vacancy-mediated diffusion mechanisms in CSAs, particularly under rugged energy landscapes caused by chemical disorders. Specifically, the ML model, trained on an extensive database of energy barriers across various local atomic environments (LAEs), serves as an accurate, on-the-fly energy barrier calculator within the kMC process. With the developed ML-kMC approach, we were able to unveil the pivotal role played by disorder-induced migration energy barriers in CSAs. Herein, our primary focus shifts to exploring the sluggish and chemically-biased diffusion mechanisms driven by interstitial diffusion in Fe-Ni CSAs, leveraging the efficiency of the ML-kMC approach. It's worth noting that interstitial-mediated diffusion fundamentally differs from vacancy-mediated diffusion, primarily due to the presence of distinct chemical signatures in interstitials. More precisely, interstitials within alloys adopt the form of dumbbells, and the type of these interstitial dumbbells exerts a significant influence on the diffusion process. Here, we have conducted an in-depth investigation into the mechanisms behind sluggish and chemically-biased diffusion by examining the various migration patterns of interstitial dumbbells. In addition, a practical AvgS-kMC based only on the mean of energy barriers with different migration patterns is proposed to efficiently and conveniently determine the diffusion coefficients mediated by interstitials in CSAs. Also, we introduce an inverse design method for optimizing sluggish diffusion properties by combining the heuristic algorithm difference evolutionary (DE) with the AvgS-kMC. The process of optimizing diffusivity highlights the critical importance of favorable migration patterns.

# Result

**Performance of the ML Model**

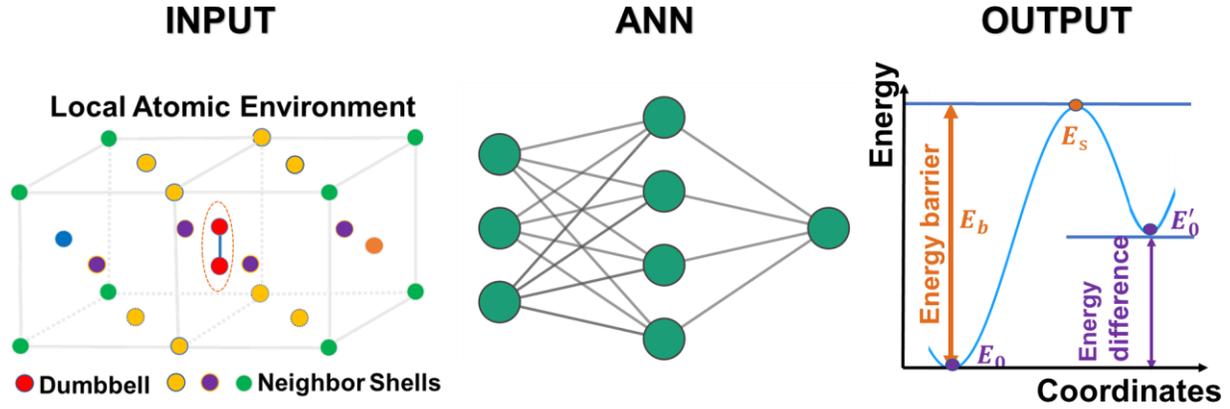

**Fig. 1** The schematic diagram of the ML method. The ANN model is trained with the input LAEs as features and output $E_b$ as the labels.

The input of the ML model was the LAEs, described by the elemental types in the nearest neighbor (NN) shells around the dumbbell interstitial, as the "INPUT" part shown in **Fig. 1**. In this work, the LAEs contain atoms within the 1$^{st}$ NN to 10$^{th}$ NN based on our test. This LAE's range is significantly larger than in vacancy case [29], consistent with the greater spatial influence of interstitial dumbbells. The output labels were $E_b$, as shown by the "OUTPUT" part in **Fig. 1**. The artificial neural network (ANN) was chosen as our ML model. In ANN, three hidden layers were employed with the node numbers (21, 19, 15) for training $E_b$. The ANN performance was evaluated by the mean absolute error (MAE) based on the 10-fold cross-validation method. The training algorithm was the Bayesian regularization backpropagation, and the epoch was 500. In **Fig. 2**, the prediction performance of the ANN model for $E_b$ achieves MAE=32.30 meV, and the Pearson correlation coefficient R=0.9609. These results indicate that the ANN model can accurately predict $E_b$ for the migration of interstitial dumbbells.

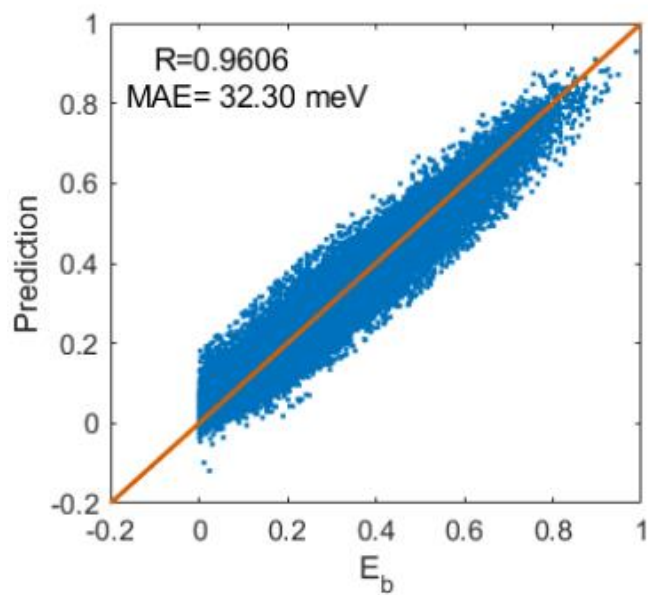

**Fig. 2** The prediction performance of the energy barrier ($E_b$). R is the Pearson correlation coefficient, and MAE is the mean absolute error.

## Diffusivity from ML-kMC

After obtaining the ML model, we combined it with the kMC method to determine the interstitial diffusivity in Fe-Ni alloys. In the kMC algorithm, the attempt frequencies for different components were determined using the method outlined by [30] at 1000K. In addition, the employed lattice constants for each component were optimized. The values of attempt frequencies and lattice constants of different components are provided in Supplementary Note 1. For each component and temperature, we conducted 600,000 kMC steps to obtain statistically meaningful diffusivity data.

The computed diffusivity with three methods under different components and temperatures are presented in **Fig. 3** as follows: the red line with circle markers represents the obtained diffusivity results of ML-kMC; the green line with square markers represents MD simulation results, which were extracted from [24] as the standard diffusivity. Additionally, to compare ML-kMC with the state-of-the-art kMC method for determining interstitial diffusion in Fe-Ni alloys, we extracted the diffusivity results from [25], represented by the pink line with star markers (Ref-kMC). From this figure, we verify that ML-kMC shows a generally better consistency with MD than Ref-kMC in most of the components under different temperatures, especially at high temperatures.

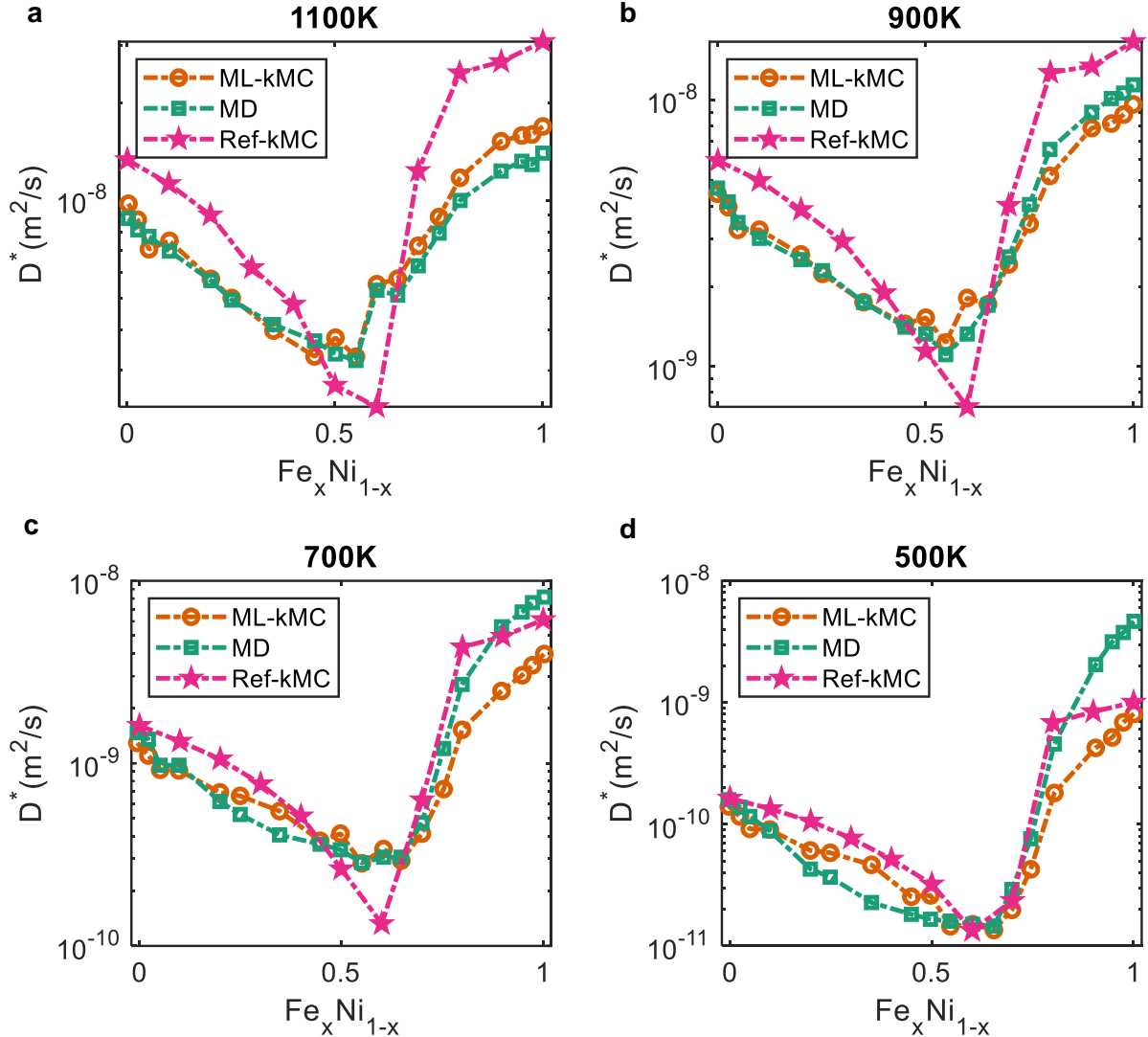

**Fig. 3** The diffusivity determined using the ML-kMC method, which is compared against the diffusivity obtained through Molecular Dynamics (MD) simulations [24] and the outcomes achieved by the state-of-the-art kinetic Monte Carlo (kMC) methodology [25]. The assessments are conducted at different temperatures: (a) 1100K, (b) 900K, (c) 700K, and (d) 500K.

In the kMC process, the energy barriers are the most crucial parameters in determining the diffusivity according to the transition state theory. The comparison between the migration energy distribution used in ML-kMC and the constant energy barriers for the eight possible dumbbell migration patterns adopted in Ref-kMC are shown in **Fig. 4a** to explain their different performances. In this figure, "Ni-Ni-Ni" denotes the initial interstitial dumbbell before migration is "Ni-Ni", and the newly formed interstitial dumbbell is also "Ni-Ni"; the "Ni" atom in the central position is the atom that actually migrates. The dot lines represent constant energy barriers, and their colors correspond to the different dumbbell migration patterns, matching the colors of the migration energy distributions. For example, the pink dot line represents the constant energy barrier of "Ni-Fe-Fe" in Ref-kMC.

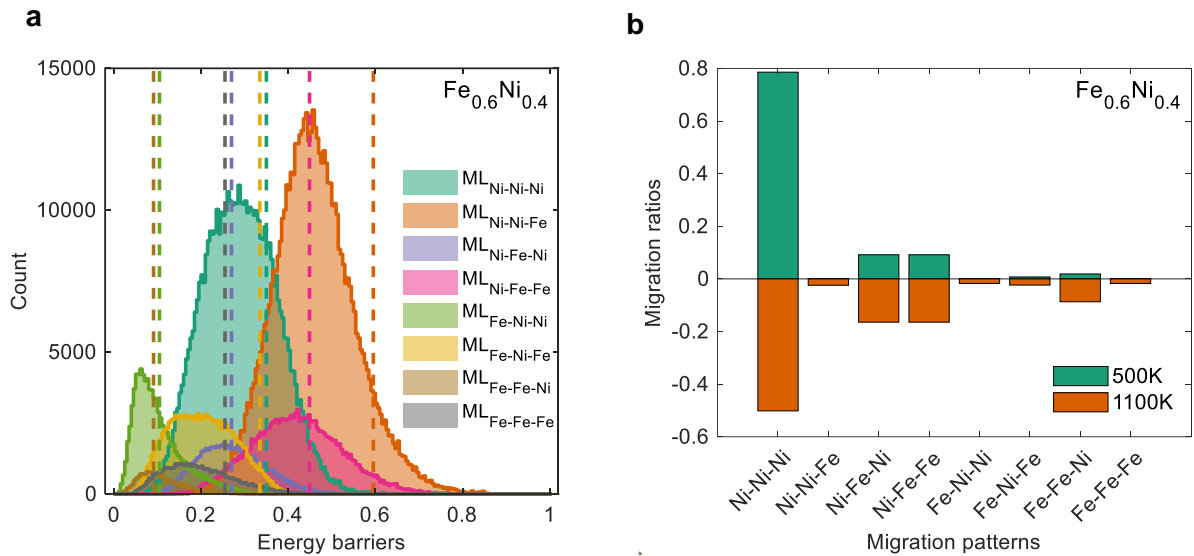

**Fig. 4** An analysis of the diffusivity difference between the ML-kMC and Ref-kMC at $Fe_{0.6}Ni_{0.4}$. (a) The energy barrier distributions for all possible migration pathways (1st NN) in ML-kMC are statistically obtained based on different migration patterns, with corresponding constant energy barriers from Ref-kMC indicated by dashed lines of the same color. (b) The ML-kMC migration ratios comparison between the low (500K) and high (1000K) temperatures, in which the migration ratios are the proportion of migration events with different migration patterns in the total migration events.

Like in MD, the ML-kMC method has the capacity to account for the influence of LAEs on interstitial migration during diffusion, whereas Ref-kMC cannot fully consider such effects. For instance, as shown in **Fig. 4a**, the impact of the same interstitial "Ni-Ni-Ni" under different LAEs yields a distribution ranging from about 0.1 eV to 0.6 eV. In contrast, in Ref-kMC, all migration energy barriers for "Ni-Ni-Ni" remain at a constant value. However, in **Fig. 4a**, it can be observed that dashed lines of the same color are generally located within the range of the corresponding energy barrier distribution with the same color. This observation suggests that utilizing constant energy barrier information can somewhat reflect the relative relationships in the actual energy spectrum. It is worth noting that for dashed lines of the same color, some of them are situated near the two extremes of the energy barrier distribution range. For instance, the orange and yellow dashed line "Ni-Ni-Fe" is located at the right end of the distribution with the same color. When the migration pathway follows this particular migration pattern, it may lead to significant deviations in the diffusion behavior.

Furthermore, the diffusivity results in **Fig. 3** reveal that Ref-kMC demonstrates a stronger alignment with ML-kMC results at lower temperatures, but this congruence diminishes as temperature increases. At lower temperatures, diffusion is predominantly driven by the migration of a single favored interstitial migration pattern, while at higher temperatures, it results from the collaborative impact migration of multiple interstitial migration patterns. For instance, the distribution of eight interstitial migration patterns obtained through the ML-kMC method for $Fe_{60}Ni_{40}$ at 500K and 1100K is depicted in **Fig. 4b**. At 500K, the primary migration involves the "Ni-Ni-Ni" interstitial type, making up around 80% of total migration, with a closely matching average energy barrier of 0.308 eV to the constant energy barrier of 0.343 eV, resulting in similar diffusion predictions. Conversely, at 1100K, the proportion of "Ni-Ni-Ni" migration decreases by approximately 30%, indicating a proportional increase in the involvement of other interstitial migration patterns. Among these interstitial migration patterns, some of them exhibit substantial disparities between the average energy barrier of their distributions and the corresponding constant energy barrier (e.g., "Ni-Ni-Fe" or "Fe-Ni-Fe"), leading to distinct diffusion predictions at high temperatures.

**Sluggish and Chemically-Biased Diffusion Mechanism**

Compared to MD, ML-kMC not only accurately considers LAEs-dependent transition energy barriers, but it also conveniently records changes in diffusion-controlling factors such as the tracer correlation factor ($f_{tr}$) and jump frequency ($\nu$). Moreover, it easily captures information on possible interstitial migration patterns and their corresponding barriers in each migration event, among other valuable information. These insights offer a comprehensive perspective for exploring the mechanisms of interstitial diffusion, including sluggish and chemically-biased diffusion. In **Fig. 5a**, $f_{tr}$ and $\nu/\nu_{0c}$ in different components are shown, in which the $\nu_{0c}$ is the attempt frequency under different components and the ratio is used to emphasize the effect of the energy barriers.

From **Fig. 5a**, we can find that the sluggish diffusion is mainly contributed by the competition between the correlation effect and jump frequency. Although $\nu/\nu_{0c}$ increases in $x \in (0, 0.45]$, $f_{tr}$ has a larger decrease magnitude in this range and leads to the total sluggish diffusivity in this component range. In addition, the strongest sluggish diffusion happens around $x \in [0.5, 0.6]$ due to the superposition decrement of both $f_{tr}$ and $\nu/\nu_{0c}$. Conversely, the significant superposition increment of both $f_{tr}$ and $\nu/\nu_{0c}$ contributes to the rapidly rising of the diffusivity with the $x$ more than 0.6, in which $\nu/\nu_{0c}$ plays a dominant role (larger variation magnitude) in weakening the sluggish diffusion effect. Therefore, the CSA design with the strongest interstitial sluggish diffusion property heavily depends on the rational composition selection where the correlation effect and jump frequency alternate control the diffusivity.

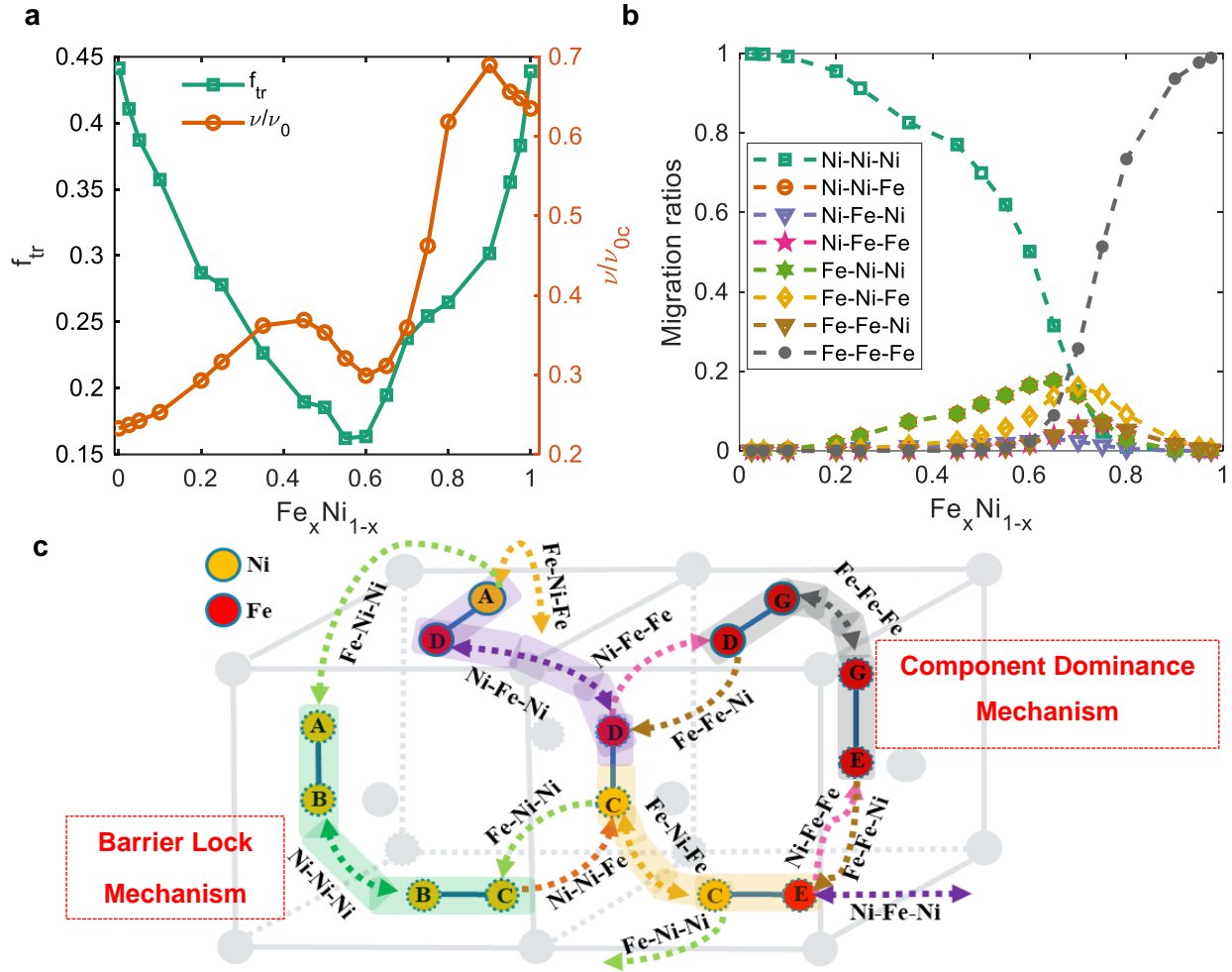

**Fig. 5** Analysis of interstitial diffusion in ML-kMC at 1100 K. (a) Tracer correlation factor ($f_{tr}$) and jump frequency ratio ($\nu/\nu_{0c}$) (b) Migration ratios of different migration patterns (c) Schematic diagram of the transformation of different interstitial migration patterns. A, B, and C are "Ni" atoms, while D, E, F, and G are "Fe" atoms; The high barrier of "Ni-Ni-Fe" and the low barrier of "Fe-Ni-Ni" locks the migration focus on the "Ni-Ni-Ni" migration pattern, forming a "Barrier Lock" mechanism; The significant rich Fe components and similar migration energy barrier of different dependent migration patterns lead to the "Fe-Fe-Fe" migration pattern dominant the migration, named as a "Component Dominance" mechanism.

In order to investigate the detailed process of chemically-bias diffusion, the statistic of migration ratios of each possible transition pattern is shown in **Fig. 5b**. With this information, we can intuitively observe the bias transition patterns and the corresponding degrees for different interstitial migration patterns. For example, the migration patterns "Ni-Ni-Ni" and "Fe-Fe-Fe" alternatively show the highest migration ratio in $x$ less than 0.7 and larger than that, respectively. This information suggests there is a "Ni-Ni-Ni" bias diffusion in $x$ less than 0.7 and a "Fe-Fe-Fe" bias diffusion in $x$ larger than that. In addition, there is a minimum bias diffusion in $x \in [0.65, 0.7]$, where the migration patterns "Ni-Ni-Ni", "Ni-Ni-Fe", "Fe-Ni-Ni", "Fe-Ni-Fe", and "Fe-Fe-Fe" has a nearly equal migration ratio. Therefore, the chemically-bias diffusion effect cannot simply be attributed to the preferred diffusion of one element type but the interstitial migration patterns.

The mutual transformation relationships between eight migration patterns are shown in **Fig. 5c**. It is worth noting that these migration patterns are not independent. For instance, the "Ni-Ni" interstitial dumbbell can migrate to form "Ni-Ni" or "Ni-Fe" but not "Fe-Ni" or "Fe-Fe". In the figure, the double-sided arrow indicates that they can mutually transform (e.g., "Ni-Ni-Ni" or "Ni-Fe-Ni"), while the one-way arrow means they can only transform in a single direction (e.g., "Ni-Ni-Fe" or "Ni-Fe-Fe"). The arrows with the same color represent the same migration events. Besides, the varying components result in a change of the LAEs around the migration interstitial types, which may greatly influence the energy barrier distribution and further lead to the significant difference in diffusion behavior on different components. Thus, In **Fig. 6**, the energy barrier distribution of each migration pattern is shown to help the analysis of the different diffusion mechanisms. Note that the distribution area of each migration pattern can intuitively reflect the proportion of each pattern among all possible migration paths.

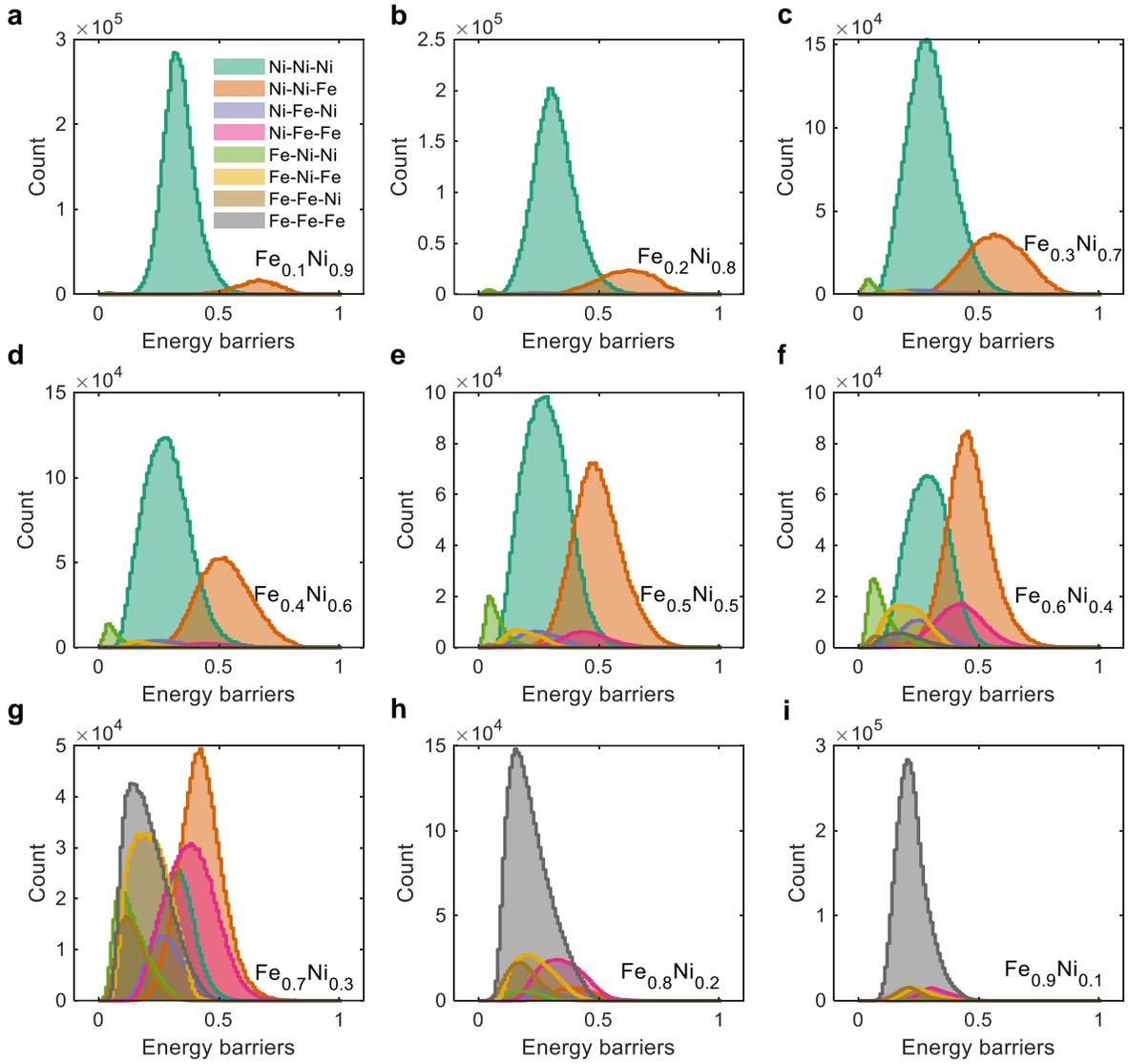

**Fig. 6** The energy barrier distributions for eight possible migration patterns in Fe$_x$Ni$_{1-x}$. (a)-(i) $x$=0.1, 0.2 to 0.9, respectively.

1) **"Barrier Lock" Mechanism.**

From **Fig. 5b**, the "Ni-Ni-Ni" migration pattern has a dominant migration ratio, which means there is a high probability that the "Ni-Ni" dumbbell selects the 1$^{st}$ NN Ni atoms to form the new "Ni-Ni" dumbbell. Interestingly, from **Fig. 6**, the energy barrier of the "Ni-Ni-Ni" migration pattern is not the smallest, which suggests that it may not have migration superior to other migration patterns. Thus, we explored how this intuitive phenomenon happens.

From **Fig. 5c**, the "Ni-Ni" interstitial dumbbell has two possible migration patterns: "Ni-Ni-Ni" and "Ni-Ni-Fe". Compare the energy barrier distribution of these two migration patterns in **Fig. 6a** to **6f**, the energy barrier distribution of "Ni-Ni-Ni" is located at a lower barrier than the "Ni-Ni-Fe" (e.g. $x$=0.1), which suggests that the "Ni-Ni-Ni" migration pattern has a much higher migration probability than the "Ni-Ni-Fe". In addition, even if the rare migration event "Ni-Ni-Fe" happened, the system would shift back to the "Ni-Ni-Ni" with very high probability due to the significantly lower energy barrier distribution (**Fig. 6**) of the "Fe-Ni-Ni" among the four possible migration patterns: "Fe-Ni-Ni", "Fe-Ni-Fe", "Ni-Fe-Ni" and "Ni-Fe-Fe" (**Fig. 5c**). Therefore, the higher barrier "Ni-Ni-Fe" and lower barrier "Fe-Ni-Ni" form a "Lock" to keep the migration pattern focused on the "Ni-Ni-Ni" migration pattern, and here we named this mechanism the "Barrier Lock". Under the function of such a mechanism, the transition shows a "Ni-Ni-Ni" bias diffusion behavior.

With the rising of the Fe component in $x \in (0, 0.6]$, the "Barrier Lock" mechanism leads to a strengthening correlation effect and increasing sluggish diffusion effect. From the analysis in **Fig. 5a**, the tracer correlation factor plays a dominant role in triggering the sluggish diffusion effect. As the Fe component increases in the "Ni-Ni" LAEs, the alternative Ni atoms around the "Ni-Ni" decreases, which raises the probability of the hop back event with the "Ni-Ni-Ni" migration pattern and contributes to the decreasing of the tracer correlation factor.

The "Barrier Lock" mechanism also contributes to explaining the concentration of highly sluggish diffusion and lowly chemically-biased diffusion that occurs when $x$>0.5. From **Fig. 6a** to **6f**, A leftward shift in the energy barrier distribution is observed with the "Ni-Ni-Fe" migration pattern. The decreasing energy barrier difference between "Ni-Ni-Ni" and "Ni-Ni-Fe" leads to a more frequent occurrence of the "Ni-Ni-Fe" migration pattern. This shifting trend weakens the "Barrier

Lock" function, allowing more migration patterns to influence the diffusion behavior. In addition, when $x>0.5$, the Fe concentration is higher than that of Ni in the LAEs of the "Ni-Ni" dumbbell, which contributes to enlarging the migration probability of the "Ni-Ni-Fe" migration pattern. As the Fe component continues to increase, the "Barrier Lock" mechanism and its weakening factors eventually reach a dynamic transition equilibrium between "Ni-Ni-Ni" and "Ni-Ni-Fe" cases. Therefore, the Fe concentrations must be more than 0.5 where the balance-off transitions can reach.

2) **"Component Dominance" Mechanism.**

From **Fig. 6g** to **6i**, there is only a slight difference in the average energy barrier among the migration patterns "Fe-Fe-Fe," "Fe-Fe-Ni," "Fe-Ni-Fe," and "Fe-Ni-Ni." However, the rich-Fe LAEs lead to the "Fe-Fe-Fe" migration pattern dominant the migrations (high migration ratios in **Fig. 5b**) during the diffusion compared to other patterns. This phenomenon is termed the "Component Dominance" mechanism, which results in "Fe-Fe-Fe"-biased diffusion when $x>0.7$. Furthermore, as the continued increase of the Fe component, the available Fe sites around the "Fe-Fe" dumbbell also increase, which would reduce the possibility of hopping back into the "Fe-Fe-Fe" migration pattern and further result in the increase of the tracer correlation factor. Besides, the left-shift energy barrier distributions of the majority of migration patterns also contribute to the increase of the jump frequency. These characteristics collectively lead to the rapid attenuation of sluggish diffusion in these components.

## The AvgS-kMC Method

From the mechanism analysis above, the energy barriers of different migration patterns are the key factor in determining the interstitial diffusion properties. Although machine learning can accurately describe this energy information, it often requires precise and extensive datasets as a foundation. The accumulation of such datasets is typically labor-intensive. Based on this consideration, here we propose a simplified model to quickly assess the interstitial diffusivity with satisfied accuracy. In this simplified model, the complex energy barrier information in kMC is simplified as the average energy barrier based on different migration patterns. This simplification can potentially facilitate the design of CSAs with the desired sluggish or chemically-bias diffusion properties.

To validate the AvgS-kMC approach, we compare the obtained diffusivity under different components with those obtained by ML-kMC as shown in **Fig. 7a**. In both kMC methods, we used the same lattice constant, initial structure, random seed, and attempt frequency, except for the energy barrier information, in which the detailed settings are provided in Supplementary Note 2. In **Figs. 7b** and **7c**, the key controlling factors of diffusivity (D*), i.e., jump frequency (ν) and correlation factor ($f_{tr}$), obtained by the two kMC methods, are shown. In **Fig. 7a**, the consistent diffusion coefficients validate the AvgS-kMC method, which indicates that AvgS-kMC could be an efficient and convenient way to evaluate diffusion coefficients for interstitial diffusion in CSAs.

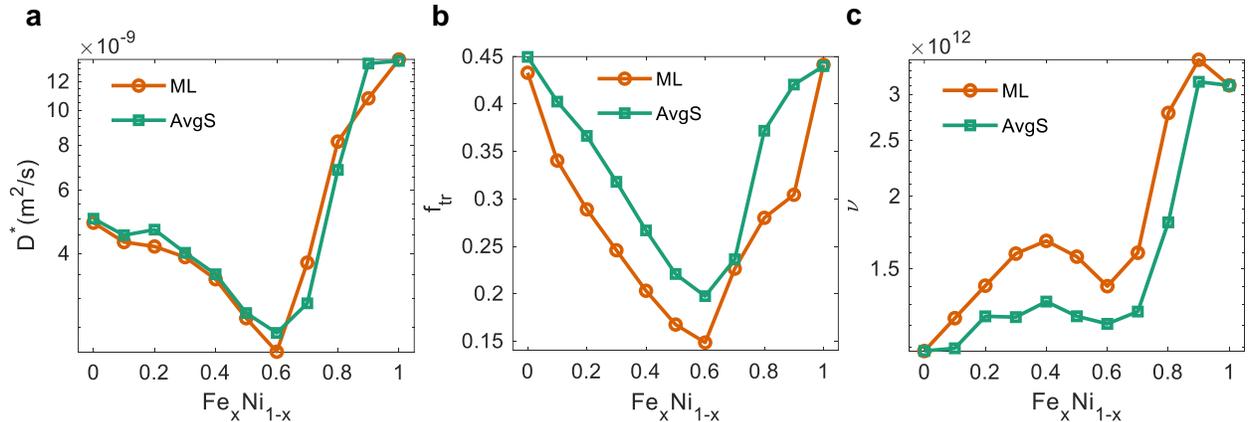

**Fig. 7** Comparison of the AvgS-kMC and ML-kMC method at 1100 K. (a) The diffusion coefficients. (b) The tracer correlation factors ($f_{tr}$), and (c) Jump frequency (ν).

The similar diffusion coefficients derived from both ML-kMC and AvgS-kMC in **Fig. 7a** can be attributed to the canceling effect resulting from the interplay between the correlation factor and the jump frequency, as illustrated in **Figs. 7b** and **7c**. In the case of averaged energy barriers $\bar{E}_b$, according to Eq. (8), the summation transforms into Zexp(-β$\bar{E}_b$), β = 1/($k_b$T). If $E_b^i$ follows a normalized distribution of $N(\bar{E}_b, \sigma)$, the sum Zexp(-β$\bar{E}_b$) will always be less than $\sum_{i=1}^{Z} \exp(-\beta E_b^i)$. Consequently, the jump frequency in the AvgS-kMC approach will consistently be lower than in ML-kMC, as **Fig. 7c** shows. On the other hand, the interstitial jumps tend to favor the lower $E_b^i$ paths. When these lower $E_b^i$ are replaced by the $\bar{E}_b$ due to the approximation of energy barrier distributions, the favorable jumping behavior diminishes and leads to a decrease in the hoping back probability of the migration atom. Thus, the tracer correlation factors are generally higher in the AvgS-kMC method for most of the components compared with ML-kMC. According to Eq. (5), the mutual compensation, where lower jump frequency and a higher tracer correlation factor exhibit similar varying ratios, results in consistency between ML-kMC and AvgS-kMC.

Considering that acquiring the energy barrier distribution for various migration patterns in CSAs is a more time-efficient and straightforward process compared to collecting a large ML training dataset, the AvgS-kMC method can serve as a rapid and reasonably accurate approach for determining diffusion coefficients resulting from interstitial diffusion.

**Inverse design for optimizing the sluggish diffusion properties in CSAs**

Given that diffusion coefficients in CSAs can be relatively accurately determined by the mean of energy barriers (MEBs) for various interstitial migration patterns with the AvgS-kMC method, we introduce an inverse design strategy to gain a comprehensive understanding of how interstitial migration patterns influence sluggish diffusion properties within CSAs. Specifically, the objective is to optimize the diffusivity by identifying the satisfying combinations of MEBs for various migration patterns. The optimization, however, adheres to certain constraints grounded in fundamental physical principles, as transformations between different interstitial migration patterns should align with general observations in Fe-Ni alloys. Firstly, the pure MEB in Fe should be lower than in Ni, implying $E_{Fe-Fe-Fe} < E_{Ni-Ni-Ni}$. Secondly, the interstitial transition Ni(Fe)-Ni(Fe)-Ni should be easier than Ni(Fe)-Ni(Fe)-Fe since the greater stability of Ni-Ni dumbbells compared to the Fe-Fe, meaning $E_{Ni(Fe)-Ni(Fe)-Ni} < E_{Ni(Fe)-Ni(Fe)-Fe}$. Consequently, the inverse design strategy transforms into a constrained optimization problem.

As the high-dimensional and constrained nature of the variable space of MEBs, tackling this optimization problem through trial and error becomes arduous. To address this challenge, we turn to the differential evolution (DE) algorithm, a well-established population-based optimizer with a track record of solving various constrained optimization problems [31-33]. The DE algorithm typically comprises three key steps: mutation, crossover, and selection. A detailed breakdown of these steps can be found in our prior work [34]. In the context of this study, the energy barriers associated with different migration patterns serve as the population within DE, while the fitness value is tailored to represent the tracer diffusion coefficient as determined by AvgS-kMC. To account for the constrained conditions on energy barriers, we incorporate a constrained violation degree (CVD) into the fitness value. Populations in DE that fall outside the boundary limitations are adjusted to conform to the boundary values. Using this optimization algorithm, we embark on a journey to minimize or maximize diffusion coefficients, all while adhering to the aforementioned constraints. Minimizing the diffusion coefficient assists in identifying the pivotal migration patterns responsible for sluggish diffusion, whereas maximizing it sheds light on the conditions under which migration patterns promote rapid diffusion. Our objective function is defined by a combination of the diffusion coefficient (D) and CVD as follows:

Minimization objective:

$$\min f(E_b^i) = D(E_b^i) \times (1 + s \times CVD), \tag{1}$$

and maximization objective:

$$\max f(E_b^i) = \frac{D(E_b^i)}{1 + s \times CVD}, \tag{2}$$

where *s* is a constant scale factor to control the penalty, and *CVD* is determined as:

$$CVD = \sum_{c=1}^{N} \max\{0, \ (E_b^{n_c} - E_b^{m_c})\}, \tag{3}$$

where $E_b^{n_c}$ and $E_b^{m_c}$ are the energy barriers of $n_c$ and $m_c$ migration patterns that have internal physical constraints, which requires $E_b^{n_c} < E_b^{m_c}$. Besides, $(E_b^{n_c} - E_b^{m_c}) < 0$ means that the current energy barrier satisfies the constraint and thus *CVD*=0. On the other hand, if $(E_b^{n_c} - E_b^{m_c}) > 0$, the value will be used as a key factor to determine the penalty; the larger $(E_b^{n_c} - E_b^{m_c})$, the stronger the penalty. Here, $c$=1, 2, …, N is the number of constraints. There is a total of five constraints: (1) $E_{Fe-Fe-Fe}<E_{Ni-Ni-Ni}$; (2) $E_{Ni-Ni-Ni}<E_{Ni-Ni-Fe}$ (3) $E_{Ni-Fe-Ni}<E_{Ni-Fe-Fe}$; (4) $E_{Fe-Ni-Ni}<E_{Fe-Ni-Fe}$; (5) $E_{Fe-Fe-Ni}<E_{Fe-Fe-Fe}$.

Throughout the DE search process, each iteration involves a population size of 30, and we set the scale factor as s=100. To expedite the evaluation of the fitness function, we harnessed parallel computation with the utilization of 30 CPU cores operating in parallel. We determined the stop criterion as the point where the objective function exhibits no further optimization over successive 100 iterations.

The optimization and the original MEBs in $Fe_{0.5}Ni_{0.5}$ at 1100K are depicted in **Fig. 8a**. In this radar plot, you can observe eight axes, each corresponding to one of the eight MEBs of interstitial migration patterns. The square brackets denote the boundary conditions we applied, based on general observations in Fe-Ni alloys, which suggest that MEBs typically fall within the range of 0.3 eV. The pink-triangle MEBs correspond to the original diffusivity (3.2814e-09 $m^2$/s); the red-circle and green-square MEBs are obtained from the minimum (6.4836e-10 $m^2$/s) and maximum (4.7550e-8 $m^2$/s) diffusivity, respectively. The center of the radar plot corresponds to 0 eV, and the maximum value of each axis is 0.75 eV for including all the varying MEBs. Comparing the original and optimized diffusion coefficients, the minimized diffusion coefficient is much lower than the original ones, whereas the maximal one shows the opposite. These results suggest that the diffusion properties can be efficiently tailored by the MEB of different interstitial migration patterns. In addition, in order to intuitively show the diffusivity effect from the MEBs, in **Fig. 8b**, the migration ratios of different migration patterns with different inverse design strategies in the kMC process are shown.

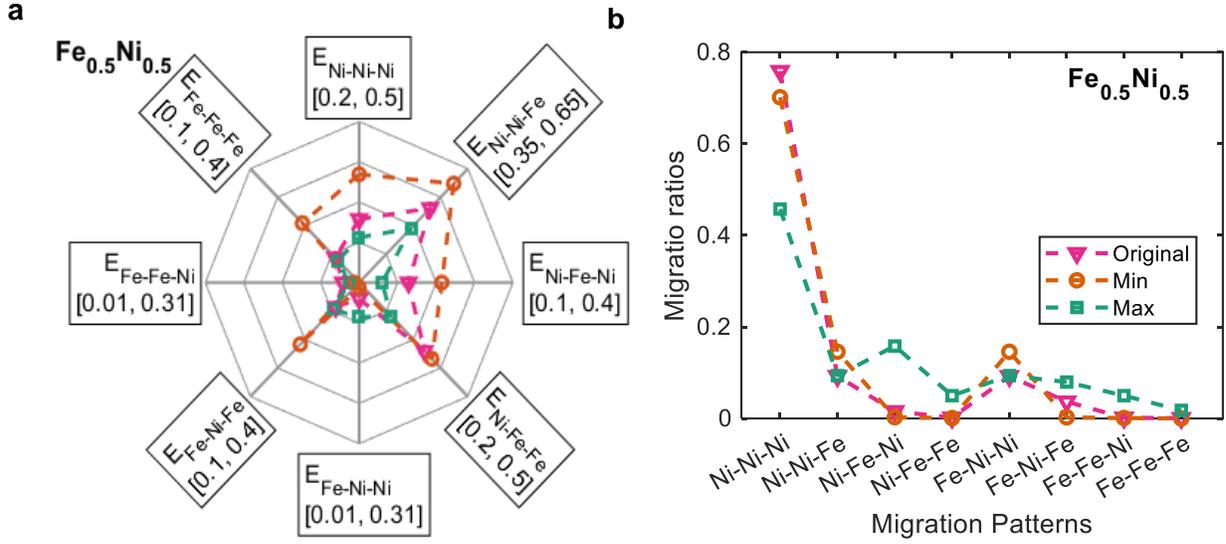

**Fig. 8** (a) The inverse design MEBs of the optimized diffusivity at $Fe_{0.5}Ni_{0.5}$. and (b) The statistics of migration ratios of different migration patterns under different inverse design strategies. The legend in (a) is the same as it in (b), and the Original, Min, and Max correspond to the original, minimum, and maximum diffusivity in inverse design strategies.

In **Fig. 8a**, when comparing the "Min" diffusivity to the "Original" one, the enhanced sluggish diffusion effect arises from two main factors. On the one hand, the elevated values of $E_{Ni-Ni-Ni}$ and $E_{Ni-Ni-Fe}$ contribute to the decrease of the jump frequency. As seen in **Fig. 8b**, the sum of the migration ratios of "Ni-Ni-Ni" (0.70) and "Ni-Ni-Fe" (0.15) migration pattern is more than 0.85 in the "Min" case, which means the migration of "Ni-Ni" dumbbell is the dominant migration event. In addition, according to Eq. (8), the larger barriers in all possible migration correspond to the smaller jump frequency. Therefore, the significantly higher values of $E_{Ni-Ni-Ni}$ and $E_{Ni-Ni-Fe}$ in the "Min" case compared to the "Original" case contribute to a significant decrease in the jump frequency. On the other hand, the lowest $E_{Fe-Ni-Ni}$ and highest $E_{Fe-Ni-Fe}$, $E_{Ni-Fe-Ni}$, $E_{Ni-Fe-Fe}$ strengthen the correlation effect. From **Fig. 5c**, the happened "Ni-Ni-Fe" migration pattern has 4 possible next migration patterns: "Fe-Ni-Ni", "Fe-Ni-Fe", "Ni-Fe-Ni", and "Ni-Fe-Fe". In the "Min" case of **Fig. 8b**, the migration ratios of "Fe-Ni-Fe", "Ni-Fe-Ni", and "Ni-Fe-Fe" are nearly zero; the migration ratios with "Ni-Ni-Fe" (about 0.15) and "Fe-Ni-Ni" (about 0.15) migration pattern are not only equal but both have considerable proportion, implying that nearly all "Ni-Ni-Fe" migration events are likely to be followed by "Fe-Ni-Ni" migration events. These phenomena significantly contribute to the reverse hopping of migration atoms, resulting in a strong correlation effect in the "Min" case. Hence, the larger migration ratios for "Ni-Ni-Fe" and "Fe-Ni-Ni" migration patterns in the "Min" case compared to the "Original" case, indicate a stronger correlation effect in the "Min" case.

In addition, the migration ratios of "Fe-Ni-Fe" also account for a non-negligible proportion in the "Original" case, which indicates that a noteworthy fraction of the happened "Ni-Ni-Fe" is not followed by the "Fe-Ni-Ni" migration pattern, impairing the reverse hopping behavior. Thus, this phenomenon contributes to weakening the correlation effect and enhancing diffusivity.

Conversely, weakening sluggish diffusion can be achieved by increasing the $E_{Fe-Ni-Ni}$ and reducing the barriers associated with various migration patterns, based on the MEBs of the "Original" case. As seen in **Fig. 8b**, the migration ratio of the higher barrier "Fe-Ni-Ni" only accounts for about 10% of the total migration events, while approximately 90% of other migration patterns have lower barriers, resulting in a significant jump frequency increase. Moreover, the enlarging of $E_{Fe-Ni-Ni}$ and shrinking of other barriers also narrow the energy barrier difference in $E_{Fe-Ni-Ni}$, $E_{Fe-Ni-Fe}$, $E_{Ni-Fe-Ni}$, and $E_{Ni-Fe-Fe}$, contributing to the weakening of correlation effect.

## Discussion

A profound understanding of sluggish and chemically-biased diffusion effects emerges from a combined analysis of migration ratios and energy barrier distribution information for various interstitial migration patterns under different compositions. Sluggish diffusion and "Ni-Ni-Ni" biased diffusion stem from the function of the "Barrier Lock" mechanism. As the continued increase in the Fe component, the energy barrier for "Ni-Ni-Fe" gradually decreases, while the energy barrier for "Fe-Ni-Ni" gradually increases. These changes weaken the effect of the "Barrier Lock" mechanism. When the effect of the " Barrier Lock " is counteracted by the influence of the increased Fe content in the atomic environment, the sluggish diffusion effect becomes most prominent, while the chemically-bias diffusion effect becomes weakest. With a further increase in the Fe component, the "Composition Dominance" mechanism leads the system toward "Fe-Fe-Fe" bias diffusion. These two controlling mechanisms may shed some light on the CSA design with the desired properties such as sluggish diffusion or chemically biased diffusion.

Inspired by the mechanism found above, we propose a more practical approach, the AvgS-kMC method, which is based on the MEBs of different migration patterns. In our comparative analysis of ML-kMC and AvgS-kMC, we emphasize the pivotal role played by the MEBs of various interstitial migration patterns in dictating diffusion behavior. In binary CSAs, there are eight possible migration patterns, and their barrier information is required to fully describe the diffusion properties. These MEBs can be determined through NEB calculations using methods such as density functional theory (DFT) or empirical potentials, although a trade-off must often be made between accuracy and computational expense. Notably, the recent development of ML interatomic potentials provides a powerful tool to reconcile these approaches. Nonetheless, it's worth noting that all available studies on the migration barriers of interstitials in CSAs [19,24,25] predominantly discuss their impact on diffusion with a focus on the constituent species rather than the specific migration patterns. Our findings in this study, however, illustrate that even for the same species, the energy barriers for various migration patterns can have contrasting effects. For instance, consider the "Ni-Ni-Fe" and "Fe-Ni-Ni" transitions, both involving Ni as the mediator. While $E_{Ni-Ni-Fe}$ contributes to decreasing jump frequency, whereas $E_{Fe-Ni-Ni}$ enhances the jump frequency. Therefore, the results here call for reporting the migration barriers for all possible interstitial migration patterns rather than just for species, as most studies do currently.

Our inverse design results by optimizing the diffusivity highlight the important role of favorable migration patterns. If a transition "lock" (e.g., Ni-Ni-Ni) is preferred, it can lead to a substantial reduction in the jump frequency, consequently significantly suppressing diffusivity. In contrast, diffusivity would be promoted if different interstitials could freely transition. These preferences for diffusion are fundamentally determined by the migration barriers associated with distinct migration patterns, which encompass information about the variations in formation energy when transitioning between different initial and final states. For instance, it is highly likely that $E_{Fe-Ni-Ni}$ is relatively low, while $E_{Ni-Ni-Fe}$ is relatively high, as Ni-Ni dumbbells are more stable than Fe-Ni. Such barriers depend on the atomic compositions experienced by the interstitial, dictating the possibility of tuning diffusion properties in CSAs.

Moreover, we also want to clarify the applicability scenarios in which these methods can be effectively applied. The ML-kMC and AvgS-kMC methods are well-suited for simulating interstitial diffusion but may not be ideal for processes involving phase transformations. This limitation arises from the inherent characteristics of the kMC method, which usually treats the lattice as a rigid structure and does not simulate atomic vibrations as the MD method does. Consequently, the kMC methods struggle to incorporate the effects of phase transformations, in contrast to MD. For instance, **Fig. 3** reveals a noticeable disparity in diffusivity between MD and ML-kMC, particularly in the range with a high Fe component at low temperatures. This difference can be attributed to the fact that in MD diffusion simulation, the initially constructed face-centered cubic (FCC) phase may transform into body-centered cubic (BCC) or other phases under the rich-Fe component and low-temperature conditions according to the Fe-Ni phase diagram, whereas the kMC-based approach focuses exclusively on the pure FCC phase for each component and temperature. However, it's important to note that this distinction does not significantly affect the core findings of this study. Our mechanistic analysis, the verification of the AvgS-kMC method, and the inverse design approach were all conducted under high-temperature conditions (1100K). At such elevated temperatures, the Fe-Ni alloy retains a pure FCC phase over a broad range of Fe compositions in the Fe-Ni phase diagram, encompassing approximately $x \in [0, 0.9]$ in $Fe_xNi_{1-x}$. Therefore, there is a high degree of consistency in diffusivity between MD and ML-kMC in such component range at high temperatures, as **Fig. 3a** shows.

## Method

The database for ML training was obtained through LAMMPS [35] based on the interatomic potential of the embedded-atom method (EAM) type developed by Bonny et al. [36]. This potential has been widely applied to simulate the defect properties in Ni-containing CSAs, and it can reproduce reasonable and consistent results with DFT calculations [19,24,37]. The initial interstitial dumbbell configuration was created by inserting an atom to form a [100] dumbbell with a lattice atom located in the center of a 10×10×10 FCC lattice. Different compositions of $Fe_xNi_{1-x}$ were considered. Note that [100] dumbbells are the most stable interstitial form in FCC Ni [37]. The migration of the [100] dumbbell to a [001] dumbbell through rotation in the {100} plane was then simulated through the NEB method [28]. In the NEB calculations, 13 intermediate images and a force convergence criterion of $1\times10^{-6}$ were used. The migration barrier $E_b$ was calculated by:

$$E_b = E_s - E_0, \qquad (4)$$

where $E_s$ is the saddle point energy, and $E_0$ is the initial state energy. In order to calculate $E_b$ under different LAEs, the atoms in the system were randomly assigned with different elemental types according to a random number generator. Specifically, the position of the inserted [100] dumbbell and the jumping position of [001] position are fixed; the elemental type at each position in the supercell changes randomly to simulate various LAEs in the dumbbell migration process.

The sequence of atomic types, including the dumbbell and atoms in LAEs, is recorded as the input features, and the migration energy barriers are target labels. This sampling method has a great advantage in dimensionality reduction. High-dimensional inputs can lead to the "curse of dimensionality" problem [38] during the training of machine learning models, resulting in poor predictive performance. Therefore, it is of utmost importance to employ dimensionality reduction methods. The interstitial energy barrier primarily depends on the types of atoms and their relative positions with respect to the dumbbell within the LAEs. Hence, the relative position plays a crucial role in determining the contribution of a specific type of atom at a given site to the energy barrier. Since the rigid-lattice kMC model features fixed atomic sites, the relative position of each atom in the LAEs remains consistent during interstitial diffusion. Consequently, we can represent each relative position in the LAEs as a one-dimensional (1D) sequence number. The input exclusively focuses on atom types, with an implicit representation of relative positions within the sequence of atom types. This transformation allows us to reduce the original 4D information of each atom in the LAEs to a 1D atom type representation.

Since the sampling method for constructing the database adopts a fixed sequence, the trained ML model cannot be directly used during kMC. As shown in **Fig. 9**, there are eight $1^{st}$ nearest neighbor (NN) atoms, which corresponds to 8 possible paths at each kMC step. To ensure consistent relative positions across these eight distinct migration cases, a coordinate transition methodology has been meticulously devised. This method serves the crucial purpose of aligning the sequence of relative positions with that of the pre-trained machine learning model, thereby facilitating accurate prediction outcomes. In this method, we build the standard coordinate system (S) according to the interstitial dumbbells that underwent the transition, namely "Initial Dumbbell" and "Next Dumbbell" in **Fig. 9**. In specific, we set vector $V_{12}$, which is the position difference of atom-1 and atom-2 at the initial state, as the x-axis. The vector $V_{32}$ is the position difference between atom-3 and atom-2 at the final state, which is set as the y-axis. The cross product of $V_{12}$ and $V_{32}$ is the z-axis, and the central position of the "Initial Dumbbell" pair is the origin. When the atom in "Initial Dumbbell" migrates to different $1^{st}$ NN atoms, we perform the same operation to build the coordinate system. Such a created S coordinate system based on the positions of the "Initial Dumbbell" and "Next Dumbbell" ensures that the relative atoms positions are consistent. Therefore, we only need to sort the atom type sequence according to the one used in model training in order to predict the migration energy barrier by the ML model.

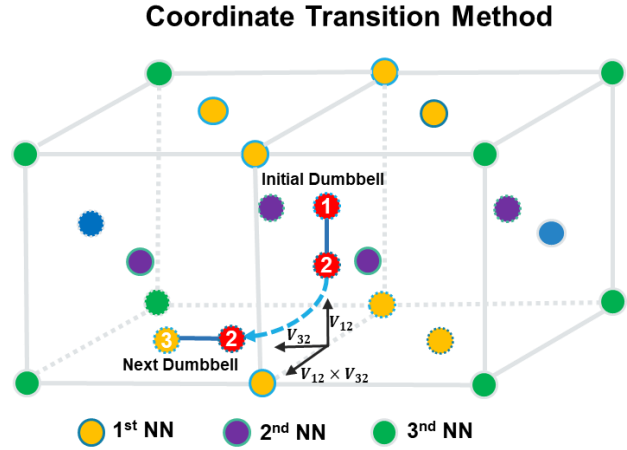

**Fig. 9** The coordinate transition method for combining the ML with the kMC in interstitial diffusion. The dumbbell interstitial migrates from the [001] plane to the [100] plane; The relative positions are determined by the vectors $V_{12}$ (difference between the initial dumbbell atom-1 and atom-2) and $V_{32}$ (difference between the next dumbbell atom-3 and atom-2). $V_{12} \times V_{32}$ represents the orthogonal vector of both $V_{12}$ and $V_{32}$ to determine the coordinate system.

In this study, the supervised ML regression model has been adopted to surrogate the complex mapping relationship between the LAEs of the dumbbell interstitial (inputting features) and $E_b$ (outputting labels). The quality of the dataset is critical to train a robust ML model that could accurately predict interstitial migration energies during migration. As the types of interstitial dumbbells (i.e., Ni-Ni, Ni-Fe, Fe-Fe) at the initial and final states have a significant influence on $E_b$, it is crucial to ensure that the dataset covers all possible transitions for different interstitial types (e.g., Ni-Ni to Ni-Fe, or Fe-Ni to Ni-Ni). In $Fe_xNi_{1-x}$, there are a total of eight possible interstitial transitions: "Ni-Ni-Ni", "Ni-Ni-Fe", "Ni-Fe-Ni", "Ni-Fe-Fe" "Fe-Ni-Ni", "Fe-Ni-Fe", "Fe-Fe-Ni", and "Fe-Fe-Fe". In this work, a total of about 70,000 migration barriers are obtained, and there are at least 500 samples for each possible transition in different concentrations.

The detailed procedure for combining ML with the kMC algorithm was presented in our previous work [29]. A total of 600,000 kMC steps are performed for each alloy composition. The tracer diffusion coefficients ($D^*$) were calculated by:

$$D^* = \frac{\langle R^2 \rangle}{6t} = \frac{r^2 f_{tr} \nu}{6}, \tag{5}$$

where $\langle R^2 \rangle$ represents the mean square displacement at time $t$; $r$ is the jump length and can be determined by $r = a/\sqrt{2}$, and $a$ is the lattice constant; $f_{tr}$ is the tracer correlation factor, which can be determined as follows:

$$f_{tr} = \lim_{n \to \infty} \frac{\sum_{i=1}^{N} R_{i,n}^2}{nr^2}, \tag{6}$$

where $n$ is the total number of jumps of the dumbbell interstitial, $N$ is the total number of atoms, and $R_{i,n}^2$ is the square displacement of the atom $i$ at the $n^{th}$ jumps. The jump frequency $\nu$ can be expressed as:

$$\nu = \frac{1}{\overline{\Delta t}}, \tag{7}$$

where $\overline{\Delta t}$ is the average time interval for each jump, and can be expressed as:

$$\overline{\Delta t} = \langle \left[ \sum_{i=1}^{Z} \nu_0 \exp\left(\frac{-E_b^i}{k_b T}\right) \right]^{-1} \rangle. \tag{8}$$

where $\nu_0$ is the attempt frequency, $E_b^i$ is the energy barrier of $i^{th}$ possible transition path, $k_b$ is the Boltzmann constant, $T$ is the temperature, and Z is the coordinate number.

## Data availability

The data that support the findings of this study are at https://github.com/Jeremy1189/interstitial-diffusion.git, or from the corresponding author upon reasonable request.

## Code availability

The codes developed for this work are available at https://github.com/Jeremy1189/interstitial-diffusion.git.

.

# Acknowledgment

This work was supported by the Research Grant Council of Hong Kong (Nos. 11200421, 8730073, 11209021, 8730065, 11214423 and C1017-21G), and the National Natural Science Foundation of China (No. 11975193). The computational resources provided by CityU Burgundy Supercomputer are highly acknowledged.